\pdfoutput=1
\documentclass{PoS}

\usepackage{amsmath}
\usepackage{amssymb}
\usepackage{cite}

\newcommand{\be}{\begin{equation}}
\newcommand{\ee}{\end{equation}}
\newcommand{\<}{\langle}
\renewcommand{\>}{\rangle}
\newcommand{\eq}{Eq.~}

\newcommand{\fig}{Fig.~}

\newcommand{\la}{\label}

\newcommand{\ba}{\begin{eqnarray}}
\newcommand{\ea}{\end{eqnarray}}

\title{Chiral dynamics in the low-temperature phase of QCD}

\ShortTitle{Chiral dynamics in the low-temperature phase of QCD}

\author{Bastian B. Brandt\\
Institut f\"ur theoretische Physik, Universit\"at Regensburg, D-93040 Regensburg, Germany}

\author{Anthony Francis, Harvey B.\ Meyer and \speaker{Daniel Robaina}\\
PRISMA Cluster of Excellence,
Institut f\"ur Kernphysik and Helmholtz~Institut~Mainz,
Johannes~Gutenberg-Universit\"at~Mainz,
D-55099 Mainz, Germany\\
        E-mail: \email{robaina@kph.uni-mainz.de}}

\abstract{We investigate the low-temperature phase of QCD and the crossover region with two light flavors of quarks.
The chiral expansion around the point $(T, m_q = 0)$ in the temperature vs. quark-mass plane indicates that a
sharp real-time excitation exists with the quantum numbers of the pion. We determine its dispersion relation
and test the applicability of the chiral expansion. The time-dependent correlators are also analyzed using the
Maximum Entropy Method (MEM), yielding consistent results. Finally, we test the predictions of ordinary
chiral perturbation theory around the point $(T = 0, m_q = 0)$ for the temperature dependence of static
observables. Around the crossover temperature, we find that all quantities considered depend only mildly on
the quark mass in the considered range 8MeV $\leq \overline{m}^{\overline{\text{MS}}} \leq$ 15MeV.}

\FullConference{The 32nd International Symposium on Lattice Field Theory\\
		 23-28 June, 2014\\
		 Columbia University New York, NY}

\begin{document}

\section{Introduction}
Lattice Gauge Theory has been widely used as a non-perturbative tool to investigate the phase diagram of QCD and the temperature dependence of key observables with the hope of understanding how the transition to the deconfined phase takes place and moreover to predict certain static and dynamic properties of the Quark-Gluon-Plasma.\\
\indent Nevertheless, in the low-temperature phase of QCD relatively little is known from the lattice about the spectral functions that encode the real-time excitations of the system \cite{HMreview}. Phenomenological models like the Hadron Resonance Gas (HRG) appear to give good estimates for static quantities like quark susceptibilities \cite{HRG}. This does not imply that the real-time excitations of the system are in any sense similar to the ordinary QCD resonances observed at zero temperature.\\
\indent In this proceedings article we want to report on our recent study addressing the question of what becomes of the pion, which is the lowest lying bound state of the theory, when it is put into a thermal bath. Specifically, our main goal will be to study its dispersion relation. The latter is no longer dictated by Lorentz invariance but takes the following form:
\be\la{eq:intro_disprel}
\omega_{\vec k}^2 = u^2(T)(m_\pi^2 + \vec k^2) + \dots
\ee
where $u(T)$ is a temperature dependent coefficient we call ``pion velocity'' and $m_\pi$ is its screening mass (see \cite{Schenk,Schenk2,Toublan,Pisarski} for previous perturbative studies). Note that $\omega_{\vec k}$ should be understood as the position of the pole in the corresponding retarded Green Function carrying pion quantum numbers. Our approach is based on the work of Son and Stephanov \cite{SonStep,SonStepL}. They propose an expansion around the point $(T, m_q=0)$ in the temperature vs. quark mass plane assuming that $T<T_{C}$. This is justified because of the Goldstone character of the pion and has the potential of increasing significantly the range of applicability with respect to the ordinary expansion around $(T=0, m_q=0)$ (see \fig\ref{fig:sketch}). In addition, we will also test the applicability of ordinary ChPT by comparing the temperature dependence of observables like the screening pion decay constant $f_\pi$ or the screening mass $m_\pi$ with their prediction in chiral perturbation theory \cite{Leutwyler}.\\
\indent The work is organized as follows: in section 2 we explain the chiral expansion and derive the most important equations for extracting the pion velocity $u(T)$. Section 3 is devoted to the Lattice setup where a brief description of our ensembles is given. In section 4 our results and conclusions are presented and finally we explain our future work in section 5. Interested readers can learn more about this work by reading our paper \cite{ourpaper}.

\section{Theory}
\subsection{Euclidean correlators in the chiral limit}
We will explain in this section how to derive the most important equations that lead to the extraction of the pion velocity $u(T)$. Ward Identities for two point functions that arise from the PCAC (partially conserved axial current) relation generalized to finite temperature guarantee that our approach is purely non-perturbative. Thus, we will consider Euclideanized QCD with two flavors of degenerate quarks on the space $S^1\times \mathbb{R}^3$, with the Matsubara cycle $S^1$ of length $\beta\equiv 1/T$. We define the vector current, axial current and the pseudoscalar density as 
\be
V_\mu^a(x) = \bar\psi  \gamma_\mu \frac{\tau^a}{2} \psi(x),
\qquad 
A_\mu^a(x) = \bar\psi  \gamma_\mu \gamma_5 \frac{\tau^a}{2} \psi(x),
\qquad P^a(x)= \bar\psi(x) \gamma_5 \frac{\tau^a}{2} \psi(x)
\ee
where $a\in\{1,2,3\}$ is an adjoint $SU(2)_{\rm isospin}$ index and $\tau^a$ is a Pauli matrix. The Dirac field is a flavor doublet $\bar{\psi}(x) = (\bar{u}(x) \bar{d}(x))$.\\
\indent Since we are looking for channels containing pion states, we choose three different euclidean correlators that couple to the pion. We recall their form and spectral representation:
\begin{eqnarray}
\delta^{ab}G_{\rm P}(x_0,\vec k) \equiv \int d^3x \; e^{-i\vec k\cdot\vec x}\; \<P^a(0) P^b(x)\> 
&=&  \delta^{ab}\int_0^\infty d\omega\, \rho_{_{\rm P}}(\omega,k)\, \frac{\cosh(\omega(\beta/2-x_0))}{\sinh(\omega\beta/2)}\,,\qquad \label{eq:PP}
\\
\delta^{ab}G_{\rm AP}(x_0,\vec k) \equiv  \int d^3x \; e^{-i\vec k\cdot\vec x}\; \<P^a(0) A_0^b(x)\> 
&=&  \delta^{ab} \int_0^\infty d\omega\, \rho_{_{\rm AP}}(\omega,k)\, \frac{\sinh(\omega(\beta/2-x_0))}{\sinh(\omega\beta/2)}\,,\qquad \label{eq:AP}
\\
\delta^{ab} G_{\rm A}(x_0,\vec k) \equiv \int d^3x\; e^{-i\vec k\cdot\vec x}\;\< A_0^a(0) A_0^b(x)\> 
&=&  \delta^{ab} \int_0^\infty d\omega\, \rho_{_{\rm A}}(\omega,k)\; 
\,\frac{\cosh(\omega(\beta/2-x_0))}{\sinh(\omega\beta/2)}\,.\qquad \label{eq:A0A0}
\end{eqnarray}
It can be shown that $G_{\rm AP}(x_0, 0)$ is fully determined by chiral Ward Identities in the limit of zero quark mass $(m=0)$ with the assumption of a non-vanishing chiral condensate, which is equivalent to $T<T_C$. It can be written as 
\be\la{eq:PA0ref}
\int d^3x\;\<P^a(0)\; A_0^b(x)\> = \delta^{ab}\frac{\<\bar\psi\psi\>}{2\beta}  \Big(x_0-\frac{\beta}{2}\Big).
\ee
This form immediately suggests that in the chiral limit at any temperature below the transition region a real-time massless excitation persists with the quantum numbers of the pion
\be\la{eq:massless_exc}
\rho_{_{\rm AP}}(\omega,0) = -\frac{\<\bar\psi\psi\>}{2} \delta(\omega).
\ee
We can also look into screening correlators (independent of $x_0$) that couple to the so called screening pion. In particular one can check via chiral WIs that the following relation is also exact in the chiral limit:
\be\la{eq:PAslab}
 \int dx_0\,d^2x_{\perp}\; \<A_3^a(x) P^b(0)\> = -\frac{\delta^{ab}}{4}\, {\rm sign}(x_3)\; \<\bar\psi\psi\>.
\ee

\subsection{Euclidean correlators at small quark mass}
\indent In virtue of \eq(\ref{eq:massless_exc}) we observe that $P(x)$ and $A_0(x)$ couple to a massless excitation when $m=k=0$, so they must also couple to a then massive excitation when a small quark mass is turned on. In this sense we expect for the case of a non-vanishing quark mass that
\be
 \int dx_0\,d^2x_{\perp}\; \<A_3^a(x) P^b(0)\> \stackrel{|x_3|\to\infty}{=}  \delta^{ab}{\rm sign}(x_3) \,c(m)\, \exp(-m_\pi |x_3|),
\ee
with $c(0)= -\frac{1}{4} \<\bar\psi\psi\>$ in view of \eq(\ref{eq:PAslab}). Now it becomes obvious that the key idea is to use the residues of the poles that we calculated in the chiral limit for the case of a small quark mass. By following this approach one can do appropriate ans\"atze for the spectral functions of \eq(\ref{eq:PP})-(\ref{eq:A0A0}) and not only restrict their form but also demonstrate that \eq(\ref{eq:intro_disprel}) is correct up to $\mathcal{O}(k^4)$ effects. We conclude that
\ba\la{eq:rhoP}
\rho_{_{\rm P}}(\omega,0) &=& -{\rm sign}(\omega) \frac{\<\bar\psi\psi\>^2\,u^2}{4 f_\pi^2}  \delta(\omega^2-\omega_{\vec 0}^2) + \dots, \\
\la{eq:rhoAP}
\rho_{_{\rm AP}}(\omega,0) &=& -\frac{\omega_{\vec 0} \<\bar\psi\psi\>}{2}\, \delta(\omega^2-\omega_{\vec 0}^2) + \dots,
\\
\la{eq:rhoA}
\rho_{_{\rm A}}(\omega,0) &=& {\rm sign}(\omega) f_\pi^2 m_\pi^2 \,\delta(\omega^2-\omega_{\vec 0}^2) + \dots
\ea
where we used the GOR relation $f^2_\pi m^2_\pi = -m \<\bar{\psi}\psi\>$ in the limit of small quark mass to introduce the screening pion decay constant $f_\pi$ which itself is defined via the following correlator
\be\la{eq:A3A3}
\int dx_0\,d^2x_{\perp} \<A_3^a(x) A_3^b(0)\> = \frac{ \delta^{ab}}{2} f_\pi^2 m_\pi e^{-m_\pi |x_3|},\qquad  |x_3|\to\infty.
\ee

\subsection{Deriving Lattice estimators for $u^2$ in terms of static quantities}
By using the spectral function of \eq(\ref{eq:rhoA}) one can easily check that
\be\la{eq:w0simple}
\omega_{\vec 0}^2 = \frac{\partial_0^2 G_{\rm A}(x_0,\vec 0)}{G_{\rm A}(x_0,\vec 0)}\Big|_{x_0=\beta/2} = 
-4m^2 \frac{G_{\rm P}(x_0,\vec 0)}{G_{\rm A}(x_0,\vec 0)}\Big|_{x_0=\beta/2}
\ee
where we used the fact that $\partial_0^2 G_{\rm A}(x_0)= -4m^2 G_{\rm P}(x_0)$. By considering the left hand side and the right hand side independently, we end up with two lattice estimators for the pion velocity $u(T)$ we call $u_f$ and $u_m$
\ba\la{eq:ush}
u_f &=& \frac{f_\pi^2 m_\pi}{2\sinh(u_f\, m_\pi \beta/2) G_{\rm A}(x_0,\vec 0)}\Big|_{x_0=\beta/2}\\
u_m &=& -\frac{4m^2}{m^2_\pi} \frac{G_{\rm P}(x_0,\vec 0)}{G_{\rm A}(x_0,\vec 0)}\Big|_{x_0=\beta/2}.
\ea
One way to test the validity of the chiral effective theory is to see whether $u_f/u_m = 1$ (see Fig. \fig\ref{fig:u}). If this is the case, it means that $G_{\rm P}(x_0)$ is dominated by the pion quasiparticle and the strength of its coupling is the one predicted by chiral WI's. It is worth nothing that at high temperatures $u_m=\mathcal{O}(m^2/T^2)$, while $u_f=\mathcal{O}(m/T)$, so that the ratio is expected to grow with temperature as it happens in practice.

\section{Lattice Setup}
\begin{figure}
\begin{center}
\includegraphics[scale=.25]{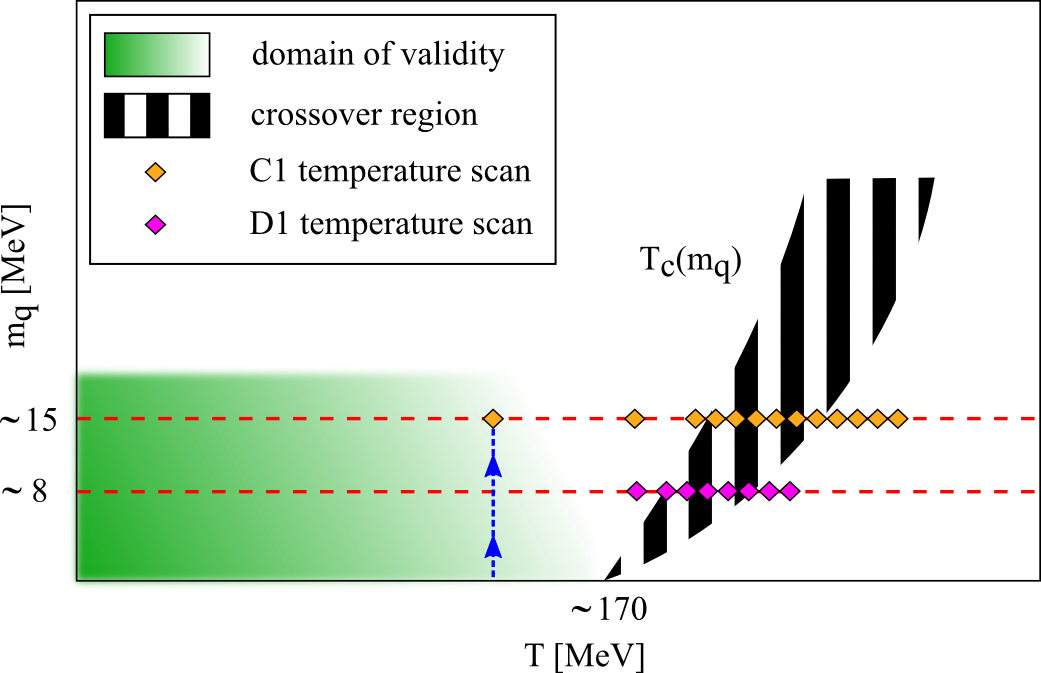}
\end{center}
\caption{Sketch of the domain of validity of the chiral effective field theory 
in the quark mass vs.\ temperature plane. 
The expansion is represented by the blue arrowed vertical line.
The quark mass on the vertical axis is understood to be $\overline{m}^{\overline{\rm MS}}$. 
The value of the critical temperature at the chiral limit $T_{c}(0) \simeq 170$ is taken from \cite{bbrandt}.}
\la{fig:sketch}
\end{figure}
In this section we describe our calculation of the relevant correlation functions that are used in the pion velocity extraction. The work is done with two temperature scans (labelled C1 and D1) at constant renormalized quark mass of $\sim 15$MeV and $\sim 8$MeV respectively (see \fig\ref{fig:sketch}). In this way our ensembles follow a line of constant physics. The gauge ensembles contain two degenerate dynamical quark flavors and we cover a temperature range $150 \le T\le 235$ MeV. For the generation of the configurations we choose the usual Wilson fermion action with a clover term for $\mathcal{O}(a)$ improvement and a non-perturbativeley determined $c_{\rm sw}$ coefficient \cite{csw}. The individual configurations are generated using the deflation accelerated DD and MP-HMC algorithms based on L\"uscher's DD-HMC package \cite{DDHMC,Luscher,Marina} on lattices of size $16\times 32^3$ with $T=1/(16a)$. The lattice scale is determined via the Sommer scale $r_0$ and its continuum value is $r_0 = 0.503(10)$fm \cite{Fritz}.\\
\indent The quantities that are needed for the analysis are: the midpoint of the axial charge correlator in the time direction $G_{\rm A}(\beta/2,\vec 0)$  as defined in \eq(\ref{eq:A0A0}); the midpoint of the pseudoscalar correlator in the time direction as defined in \eq(\ref{eq:PP}); the screening pion mass $m_\pi$ which we get from fitting the exponent of the pseudoscalar correlator along one spatial direction and the screening pion decay constant $f_\pi$ which we obtain by fitting the amplitude of \eq(\ref{eq:A3A3}). The quarkmass $m$ is renormalized in the $\overline{\rm MS}$ scheme at a scale of $\mu=2$GeV.\\
\indent In addition to scans C1 and D1 we additionally use a zero-temperature CLS test-ensemble labelled A5 with the same bare parameters as the lowest temperature ensemble of C1 at a pion mass of 290MeV but a time extent of $64a$ instead of $16a$. This enables us to test ordinary ChPT temperature predictions for $m_\pi$ and $f_\pi$ and gives an important check for our lattice estimators $u_f$ and $u_m$ since we expect that $\lim_{T \to 0} u(T) = 1$.

\section{Results \& Conclusions}
We begin by quoting the results we obtained for the pion velocity for the zero-temperature test-ensemble A5:
\be
u_f = 0.96(2) \qquad u_m = 0.92(6) \qquad u_f/u_m = 1.04(4).
\ee
This adds to our confidence that the estimators work as expected and that the chiral expansion is applicable. In \fig\ref{fig:u} we see the result for both estimators $u_f$ and $u_m$ at $\overline{m}^{\overline{\rm MS}} = 15$MeV (C1 scan). The ratio gives us information whether we are in the domain of validity region or not (see \fig\ref{fig:sketch}); and we observe that above 190 MeV we enter the crossover region and the interpretation of $u$ as the pion velocity becomes unreliable. Nevertheless, it appears likely that $u_f(T\simeq 150$MeV) = 0.88(2) indeed provides a valid estimate for the pion quasiparticle velocity. This indicates a violation of boost invariance through the presence of the medium.\\
\begin{figure}
\begin{center}
    \includegraphics[width=.48\textwidth]{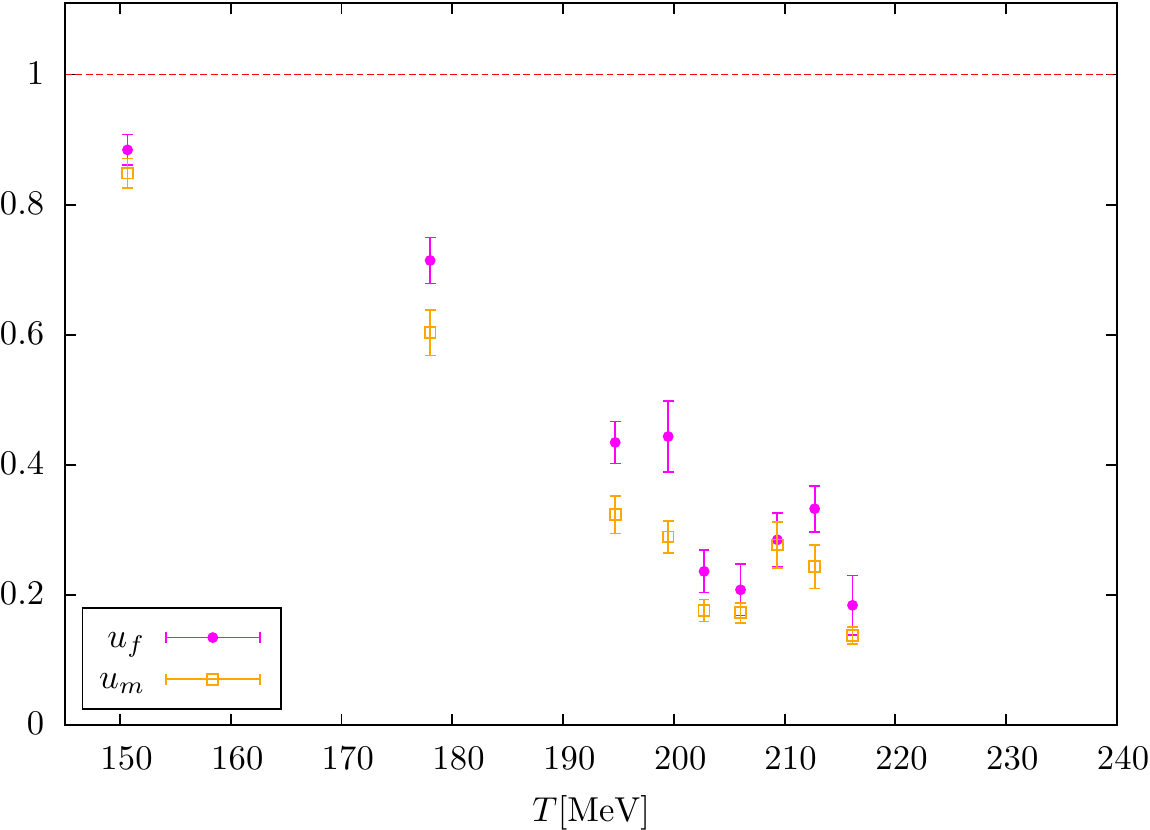}
    \includegraphics[width=.48\textwidth]{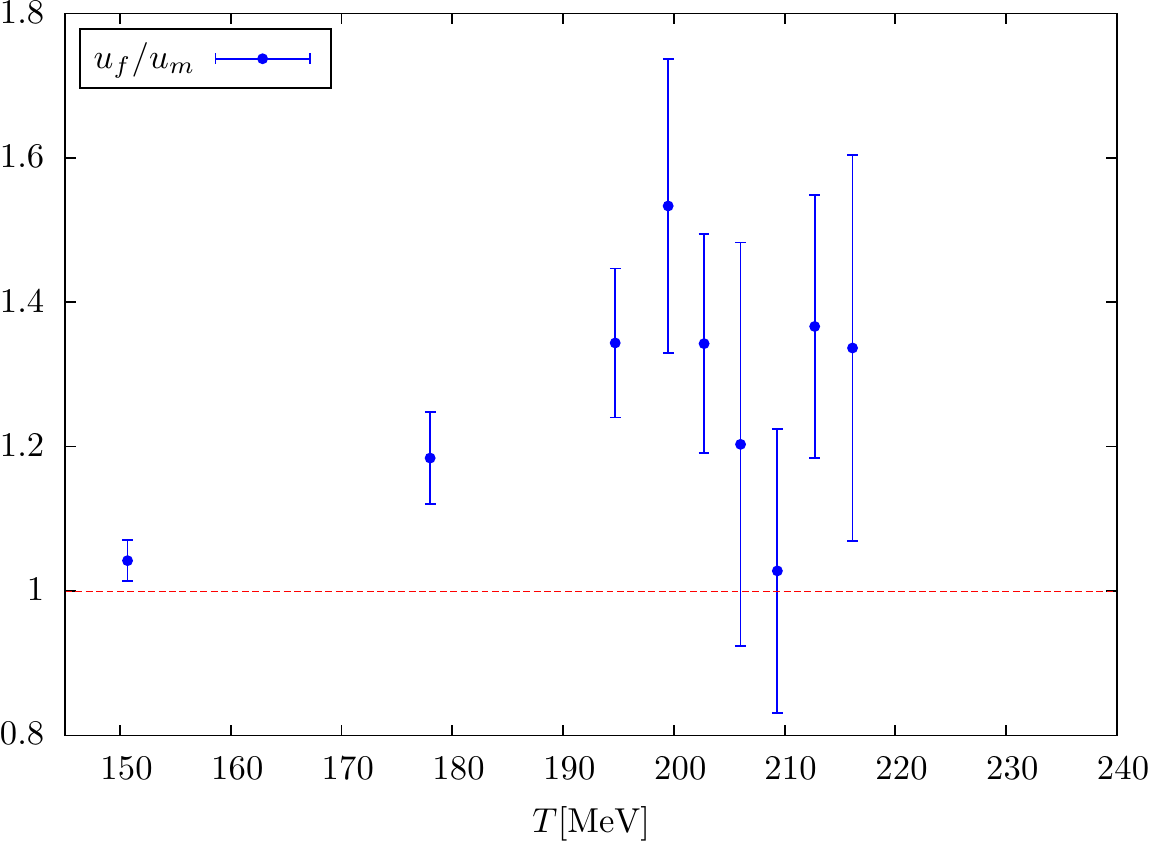}
\end{center}
\caption{Left: The two estimators of the pion velocity in the C1 scan. Right: Ratio of the estimators, which serves as a test of the chiral prediction.}
\la{fig:u}
\end{figure}
\indent In view of \fig\ref{fig:mfc} we see that at $T\simeq 150$MeV the screening pion mass $m_\pi$ and its associated pion decay constant $f_\pi$ have changed only about 5\% from their $T=0$ values. This is in agreement with the chiral expansion around $(T=0, m_q=0)$ for which we display both at 1-loop level accuracy the infinite volume system and our particular aspect ratio of $LT=2$ curve. We conclude that for temperatures $T \gtrsim 170$MeV the behavior of the screening quantities is no longer described by ChPT.\\
\indent In addition to this ``static'' analysis we also reconstructed the spectral functions of \eq(\ref{eq:PP}) and \eq(\ref{eq:A0A0}) using a Maximum Entropy Method (MEM). It can be easily shown that the area under the curve $\rho_{\rm A}$ is related to the pion velocity
\be
\mathcal{A}(\Lambda) = 2\int^{\Lambda}_{0} \frac{d\omega}{\omega}\rho_{\rm A}(\omega) = \frac{f^2_\pi}{u^2}.
\ee\\  
\indent After performing a systematic study of the $\Lambda$ dependence, we find good agreement between the MEM based results and the results for $(f_\pi/u_f)^2$. In the lowest-temperature ensemble of the C1 scan, MEM data overshoots the static results. A possible reason is that the ansatz for $\rho_{\rm A}(\omega)$ (see \eq(\ref{eq:rhoA})) is a single `delta function', while the result of the MEM reconstruction exhibits a more complicated spectral weight distribution. We observe however that the agreement improves if one uses the estimator $u_m$ instead of $u_f$ when comparing to $(f_\pi/u)^2$. In addition, cutoff effects could still be affecting the results at this lattice spacing. One should keep in mind that MEM extracts this quantity from the area under the pion peak while the other method uses its position on the $\omega$-axis.\\
\begin{figure}[h]
\includegraphics[width=.48\textwidth]{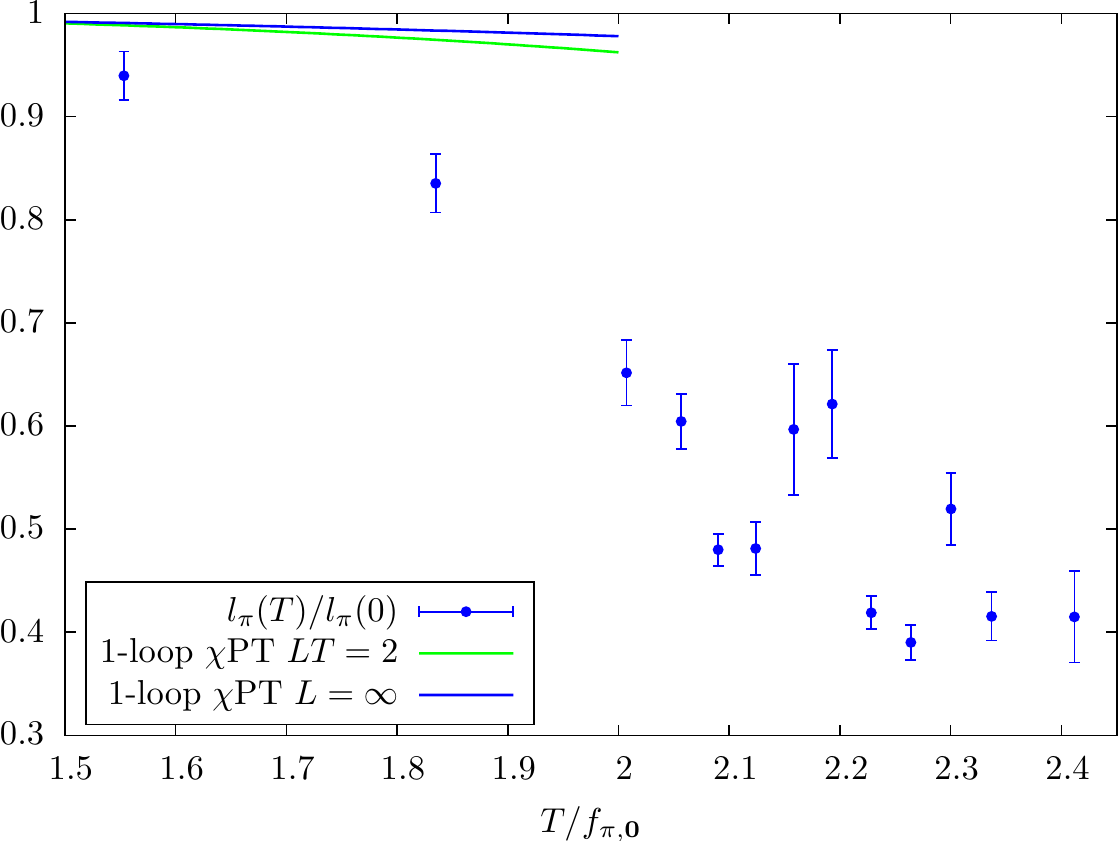}
\includegraphics[width=.48\textwidth]{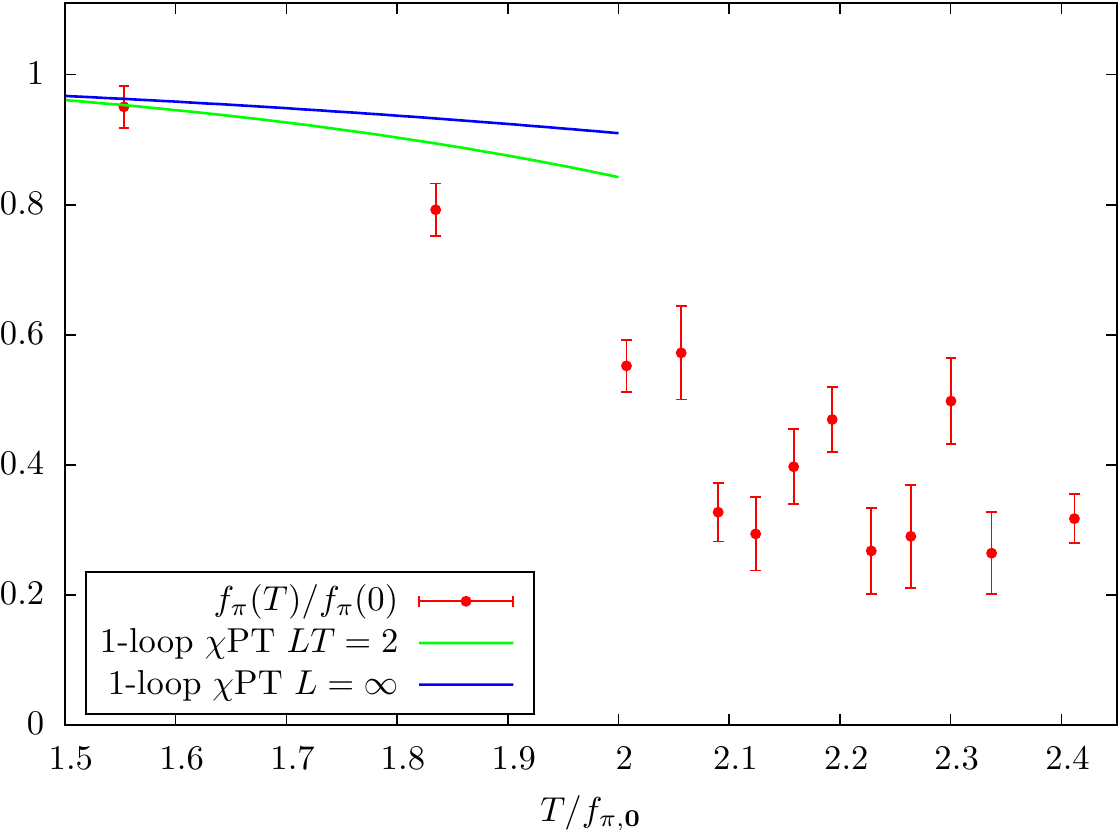}
\caption{Inverse screening mass $l_\pi\equiv m_\pi^{-1}$ (left) and 
screening pion `decay constant' (right) in the C1 scan, divided 
by the same quantity at $T\simeq0$ extracted from the A5 ensemble. 
The displayed error bars represent the statistical errors originating from 
the ensembles of the C1 scan and from ensemble A5. The temperature is normalized with the ``time''-pion decay constant extracted from $G_{\rm A}(x_0)$ for A5 which takes a value of $f_{\pi, {\bf 0}}=97(3)$MeV.}
\la{fig:mfc}
\end{figure}

\section{Outlook}
All results presented here are still affected by cutoff effects and finite volume corrections. The set of ensembles is now being extended to bigger aspect ratios of $LT=3$ and 4 in order to check for finite-volume effects. For further testing the functional form of the dispersion relation of \eq(\ref{eq:intro_disprel}), it would be very interesting to perform a similar analysis at nonzero momentum $\vec{k}$. In order to probe the $T=0$ chiral effective theory, additional simulations between 100-150MeV are desirable. It is also worth studying how far up in the quark mass the GOR relation holds at finite temperature.\\
\indent Our current study has a non-physical quark content and is done with still relatively heavy quarks. To make contact with nature one might think about simulations with physical light quark masses and a dynamical strange quark.

\section*{Acknowledgments}
We are grateful for the access to the zero-temperature ensemble used here, made available to us through CLS. We acknowledge the use of computing time for the generation of the gauge configurations on the JUGENE and JUROPA supercomputers of the Gauss Centre for Supercomputing located at Forschungszentrum J\"ulich, Germany; the finite-temperature ensemble was generated within the John von Neumann Institute for Computing (NIC) project HMZ21. The correlation functions were computed on the dedicated QCD platform ``Wilson'' at the Institute for Nuclear Physics, University of Mainz. This work was supported by the \emph{Center for Computational Sciences in Mainz} as part of the Rhineland-Palatinate Research Initiative and by the DFG grant ME 3622/2-1 \emph{Static and dynamic properties of QCD at finite temperature}.

\end{document}